\documentclass[twocolumn,prd]{revtex4-1}
\usepackage{amsmath}
\usepackage[compat=1.1.0]{tikz-feynman}
\numberwithin{equation}{section}
\begin{document}
\begin{abstract}
In non-supersymmetric covariant quantum gravity theory, for each system of gravity coupled with single field is one-loop divergent. 
Since adding other fields or other interactions to each system generates more possible counter-Lagrangian terms, there is room for improvement to restore renormalizability.
In this paper, we consider Einstein-Maxwell fields coupled with electrically charged scalar which is the simplest model among the systems of gravity coupled with multiple fields having their own interaction.
First, we introduce how to calculate the possible one-loop diagrams in Einstein-SQED system and show that this system is non-renormalizable. 
\end{abstract}
\title{One-loop divergences of quantum gravity\\coupled with scalar electrodynamics}
\date{10 December 2017}
\author{Hyun Ju Go}
\affiliation{Chung-Ang University} 
\newcommand{\Lagr}{\mathcal{L}}
\newcommand{\cuN}{\mathcal{N}}
\newcommand{\cuT}{\mathcal{T}}
\newcommand{\cuQ}{\mathcal{V}}
\newcommand{\cuA}{\mathcal{W}}
\newcommand{\cuH}{\mathcal{H}}
\newcommand{\cuI}{\mathcal{I}}
\newcommand{\cuJ}{\mathcal{J}}
\maketitle

\section{Introduction}
The quantum field theory of gravitation have been developed from Feynman's pioneer works \cite{Feynman_1}. 
Feynman showed that the self-consistent spin-2 quantum field theory is Einstein's general relativity. 
Therefore, Einstein-Hilbert action acts as suitable action for the quantum gravity. 
From this action, it is possible to calculate every tree-level diagrams by elementary methods. 
Furthermore he tried to attack one-loop diagrams and suggested fictitious quanta for unitarity of S-matrix.

After Feynman's works, Bryce DeWitt developed Feynman's results \cite{DeWitt_1,DeWitt_2}. 
He formulated manifestly covariant quantum gravity using background field method. 
From this formulation, tree theorem was proved and the algorithm for S-matrix calculations containing arbitrary order radiative corrections was derived. 
In this algorithm, the fictitious quanta for arbitrary order was introduced. 
DeWitt also analyzed non-renormalizability of quantum gravity by conventional power counting method and presented tentative proposals for dealing with this situation.

An algorithm for counter-Lagrangian of one-loop diagram was introduced by G. 't Hooft \cite{Hooft1} and this algorithm extended to include gravitation \cite{Hooft2}.
Applying this algorithm, one-loop divergences of quantum gravity coupled with scalar fields, vector fields or Yang-Mills fields were proved explicitly \cite{Hooft2,Deser_1,Deser_2}. 
For fermionic field, the situation is quite different. Firstly, one can't use metric fields as gravitational variables. 
Instead of this, fermionic field has to interact with vierbein field. 
Furthermore, t'Hooft algorithm isn't applicable for this case because of the form of Lagrangian. 
S. Deser and P. van Nieuwenhuizen solved this problem by explicit calculation of the diagrams with eight external fermions and showed that Einstein-Dirac system is also non-renormalizable \cite{Deser_3}. 

In this paper, we consider Einstein-Maxwell fields coupled with electrically charged scalar which is the simplest model among the systems of gravity coupled with multiple fields having their own interaction.
First, we calculate the possible one-loop diagrams in Einstein-SQED system and show that this system is non-renormalizable.

The rest of this paper is organized as follows. 
In section 2, the Lagrangian for one-loop diagrams of the Einstein-Maxwell fields coupled with electrically charged scalar is obtained using background field method.
In section 3, the Lagrangian for one-loop diagrams is transformed into more elegant form and Feynman rules for one-loop diagrams are derived.
Finally, in section 4, the non-renormalizability of Einstein-SQED system is showed using already known results and equation of motion.

\section{The background field method for one-loop diagrams}
We start with gravitational field $\bar g_{\mu\nu} $, scalar field $\bar\varphi $ and electromagnetic potential $\bar A_\mu $. 
From these variables, the Lagrangian for Einstein-Maxwell fields coupled with electrically charged scalar is
\begin{equation}
\Lagr=-(-\bar g)^{1/2}(\bar R + (D_\mu\bar\varphi)^*\bar g^{\mu\nu}D_\nu\bar\varphi+\frac1 4\bar F_{\mu\nu}\bar F_{\alpha\beta}\bar g^{\mu\alpha}\bar g^{\nu\beta}) \label{1}
\end{equation} 
where, $\sqrt{\bar g}$=(det $\bar g_{\mu\nu})^{1/2}$, $\bar R$ is the scalar curvature, $D_\mu\bar\varphi=\partial_\mu\bar\varphi-i\bar A_\mu\bar\varphi$ 
and $\bar F_{\mu\nu} \equiv \partial_\mu\bar A_\nu-\partial_\nu\bar A_\mu$.

The fields $(\bar g_{\mu\nu},\bar\varphi,\bar A_\mu)$ are splitted into background fields $(g_{\mu\nu},\tilde\varphi,A_\mu)$ plus quantum fields $(h_{\mu\nu},\varphi,a_\mu)$ 
to apply background field method.
Then the equation of motion and one-loop amplitudes are calculated by expanding (\ref{1}) various functions of field variables with respect to quantum fields up to second order. 
For the scalar curvature and field strength tensor, calculation results can be found in many literatures such as \cite{Hooft2,Deser_1}.
The Interaction Lagangian in Einstein-SQED up to 2nd order is
\begin{equation}
\begin{split}
\Lagr_I{}&\equiv\sqrt{\bar g}\bar g^{\mu\nu}D_\mu{\bar\varphi}^*D_\nu\bar\varphi=\\
{}&(1+\frac1 2h^\alpha_\alpha-\frac1 4h^\alpha_\beta h^\beta_\alpha+\frac1 8(h^\alpha_\alpha)^2)(g^{\mu\nu}-h^{\mu\nu}+h^\mu_\alpha h^{\alpha\nu})\\
{}&((\tilde D_\mu\tilde\varphi)^*+(\tilde D_\mu\varphi)^*+ia_\mu{\tilde\varphi}^*+ia_\mu\varphi^*)\\
{}&((\tilde D_\nu\tilde\varphi)+(\tilde D_\nu\varphi)-ia_\nu{\tilde\varphi}-ia_\nu\varphi) \label{2}
\end{split}
\end{equation}
here, we define $\tilde D_\mu=\partial_\mu-iA_\mu$ to distinct $\bar D_\mu$.
By including the result of scalar curvature and field strength tensor, the $\Lagr_2$  can be expanded by collecting the terms containing any two quantum fields as follows,
\begin{widetext}
\begin{equation}
\begin{split}
\Lagr_2={}&(-g)^{1/2}[-\frac1 2(D_\nu h_{\alpha\beta})P^{\alpha\beta\rho\sigma}(D^\nu h_{\rho\sigma})+\frac1 2(h_\mu-\frac1 2D_\mu h)^2
-\frac1 2(D_\nu a_\mu)^2+\frac1 2(D_\mu a_\nu)(D^\nu a^\mu)\\
{}&-{(\partial_\nu\varphi)}^*\partial^\nu\varphi
-\varphi^*A_\nu A^\nu\varphi+i\partial_\mu\varphi^*A^\mu\varphi-i\partial_\mu\varphi A^\mu\varphi^*\\
{}&+\frac1 2h_{\alpha\beta}(X_g+X_e+X_s)^{\alpha\beta\rho\sigma}h_{\rho\sigma}+h_{\alpha\beta}Q^{\alpha\beta\rho\sigma}D_\rho a_\sigma
-a_\mu (g^{\mu\nu}{\tilde\varphi}^*\tilde\varphi)a_\nu\\
{}&+h_{\alpha\beta}B^{\alpha\beta\rho}\partial_\rho\varphi^*+ih_{\alpha\beta}B^{\alpha\beta\rho}A_\rho\varphi^*
+h_{\alpha\beta}C^{\alpha\beta\rho}\partial_\rho\varphi+ih_{\alpha\beta}C^{\alpha\beta\rho}A_\rho\varphi+ih_{\alpha\beta}(B^{\alpha\beta\rho}-C^{\alpha\beta\rho})a_\rho\\
{}&-i{(\partial_\mu\varphi)}^*(g^{\mu\nu}\tilde\varphi) a_\nu+i\partial_\mu\varphi(g^{\mu\nu}{\tilde\varphi}^*) a_\nu
+ia_\nu g^{\mu\nu}(\tilde D_\mu\tilde\varphi-iA_\mu \tilde\varphi)\varphi^*-ia_\nu g^{\mu\nu}(\tilde D_\mu\tilde\varphi^*+iA_\mu \tilde\varphi^*)\varphi]
\end{split} \label{3}
\end{equation} 
\end{widetext}
and symbols for gravitational fields in the equation are calculated from the expansion as listed in below,
\begin{subequations}
\begin{align}
P^{\alpha\beta\rho\sigma}{}&=\frac1 2g^{\alpha\rho}g^{\beta\sigma}-\frac1 4g^{\alpha\beta}g^{\rho\sigma}\\
{X_g}^{\alpha\beta\rho\sigma}{}&=P^{\alpha\beta\rho\sigma}R-g^{\alpha\rho}R^{\beta\sigma}+g^{\alpha\beta}R^{\rho\sigma}+R^{\alpha\rho\beta\sigma}\\
{X_e}^{\alpha\beta\rho\sigma}{}&=P^{\alpha\beta\rho\sigma}\frac1 4F^2-\frac1 2F^{\alpha\rho}F^{\beta\sigma}-g^{\alpha\rho}F_2^{\beta\sigma}+\frac1 2g^{\alpha\beta}F_2^{\rho\sigma}\\
\begin{split}
{X_s}^{\alpha\beta\rho\sigma}{}&=-2g^{\sigma\beta}{(\tilde D^\rho\tilde\varphi)}^*(\tilde D^\alpha\tilde\varphi)+g^{\alpha\beta}{(\tilde D^\rho\tilde\varphi)}^*(\tilde D^\sigma\tilde\varphi)\\
{}&-\frac1 4g^{\alpha\beta}g^{\rho\sigma}{(\tilde D^\nu\tilde\varphi)}^*(\tilde D^\nu\tilde\varphi)+\frac1 2g^{\sigma\beta}g^{\rho\alpha}{(\tilde D^\nu\tilde\varphi)}^*(\tilde D^\nu\tilde\varphi) \label{4}
\end{split}
\end{align}
\end{subequations}
similarly, symbols for gravitational field coupled to Maxwell field or scalar field in the equation are
\begin{subequations}
\begin{align}
Q^{\alpha\beta\rho\sigma}={}&2g^{\alpha\rho}F^{\beta\sigma}-\frac1 2g^{\alpha\beta}F^{\rho\sigma}\\
B^{\alpha\beta\rho}={}&-\frac1 2g^{\alpha\beta}\tilde D^\rho\tilde\varphi+g^{\rho\alpha}\tilde D^\beta\tilde\varphi\\
C^{\alpha\beta\rho}={}&-\frac1 2g^{\alpha\beta}{(\tilde D^\rho\tilde\varphi)}^*+g^{\rho\alpha}{(\tilde D^\beta\tilde\varphi)}^*\label{5}
\end{align}
\end{subequations}
where ${F_2}^{\mu\nu}\equiv{F_2}^{\nu\mu}\equiv F^\mu_\alpha F^{\nu\alpha}$ and ${F_2}^\mu_\mu\equiv F^2$.

On the other hand, quadratic part of our Lagrangian is modified to obtain the Feynman rules for S-matrix. 
First, consider the following gauge transformations,
\begin{subequations}
\begin{align}
\begin{split}
h'_{\mu\nu}={}&h_{\mu\nu}+(g_{\mu\alpha}D_\nu+g_{\nu\alpha}D_\mu)\eta^\alpha+\kappa[(h_{\mu\alpha}D_\nu\\
{}&+h_{\nu\alpha}D_\mu)\eta^\alpha+\eta^\alpha D_\alpha h^{\mu\nu}] \label{6.1}
\end{split}\\
a'_\mu={}&a_\mu+\eta^\alpha F_{\alpha\mu}+D_\mu\eta^5+\kappa(a_\alpha D_\mu\eta^\alpha+\eta^\alpha D_\alpha a_\mu) \label{6.2}
\end{align}
\end{subequations}
Our original action is then invariant under these transformations,
\begin{equation}
\int d^4x'\Lagr({\bar g}',{\bar A}',{\bar\varphi}',{({\bar\varphi}^*)}')=\int d^4x\Lagr(\bar g,\bar A,\bar\varphi,{\bar\varphi}^*) \label{7}
\end{equation}
To obtain Feynman rules, it is needed to choose gauge fixing terms $-\frac1 2{C_\mu}^2$ for gravitational fields and vector fields respectively and include ghost Lagrangian in our calculations. From the form of (\ref{3}), 
one can choose $C_\mu$ as follows,
\begin{subequations}
\begin{align}
C_a={}&(-g)^{\frac1 4}e_\alpha^\mu(h_\mu-\frac1 2D_\mu h) \label{8.1}\\
C_5={}&(-g)^{\frac1 4}D_\mu a^\mu \label{8.2}
\end{align}
\end{subequations} 
where $e_\alpha^\mu$ is a square root of a metric field which is called a vierbein field.
With these gauge fixing terms, we can finally write quadratic Lagrangian for non-ghost parts :
\begin{widetext}
\begin{equation}
\begin{split}
\Lagr_{NG}={}&(-\bar g)^{1/2}[-\frac1 2(D_\nu h_{\alpha\beta})P^{\alpha\beta\rho\sigma}(D^\nu h_{\rho\sigma})-\frac1 2(D_\nu a_\mu)^2-{(\partial_\nu\varphi)}^*\partial^\nu\varphi
-\varphi^*A_\nu A^\nu\varphi+i\partial_\mu\varphi^*A^\mu\varphi-i\partial_\mu\varphi A^\mu\varphi^*\\
{}&+\frac1 2h_{\alpha\beta}(X_g+X_e+X_s)^{\alpha\beta\rho\sigma}h_{\rho\sigma}+h_{\alpha\beta}Q^{\alpha\beta\rho\sigma}D_\rho a_\sigma
-a_\mu (-\frac1 2R^{\mu\nu}+g^{\mu\nu}{\tilde\varphi}^*\tilde\varphi)a_\nu\\
{}&+h_{\alpha\beta}B^{\alpha\beta\rho}\partial_\rho\varphi^*+ih_{\alpha\beta}B^{\alpha\beta\rho}A_\rho\varphi^*
+h_{\alpha\beta}C^{\alpha\beta\rho}\partial_\rho\varphi+ih_{\alpha\beta}C^{\alpha\beta\rho}A_\rho\varphi+ih_{\alpha\beta}(B^{\alpha\beta\rho}-C^{\alpha\beta\rho})a_\rho\\
{}&-i{(\partial_\mu\varphi)}^*(g^{\mu\nu}\tilde\varphi) a_\nu+i\partial_\mu\varphi(g^{\mu\nu}{\tilde\varphi}^*) a_\nu
+ia_\nu g^{\mu\nu}(\tilde D_\mu\tilde\varphi-iA_\mu \tilde\varphi)\varphi^*-ia_\nu g^{\mu\nu}(\tilde D_\mu\tilde\varphi^*+iA_\mu \tilde\varphi^*)\varphi]
\label{9}
\end{split}
\end{equation}
\end{widetext}
Here, the Ricci identity is used
\begin{equation}
\begin{split}
(D_\alpha D_\beta-D_\beta D_\alpha)A^\mu={}&R^\mu_{\gamma\alpha\beta}A^\gamma\\
(D_\mu D_\beta-D_\beta D_\mu)A^\mu={}&-R_{\mu\beta}A^\mu\label{10}
\end{split}
\end{equation}
On the other hand, the ghost Lagragian $\Lagr_G$ can be calculated by subjecting $C_\mu$ to the gauge transformations (\ref{6.1}),(\ref{6.2}). 
From (\ref{8.1}) and (\ref{8.2}), we find
\begin{equation}
\Lagr_G=(-g)^{\frac1 4}
(\phi^{*\alpha},\chi^*)
\begin{pmatrix}
e_{\alpha\beta}D_\nu D^\nu-R_{\alpha\beta} & 0\\
-(D^\lambda F_{\lambda\beta})-F_{\lambda\beta}D^\lambda & D_\nu D^\nu
\end{pmatrix}
\begin{pmatrix}
\phi^\beta \\
\chi 
\end{pmatrix}\label{11}
\end{equation}
where, $\phi^\alpha$ is a vector ghost and $\chi$ is a scalar ghost.

\section{Feynman rules}
In this section, our Lagrangian is transformed into more elegant form and Feynman rules are derived.  
Let us consider the following form of Lagrangian,
\begin{equation}
\Lagr=(-g)^{1/2}(\phi^*_iD_\mu W^{\mu\nu}_{ij} D_\nu \phi_i+2\phi^*_iN^{\mu}_{ij}\partial_\mu\phi_j+\phi^*_iM_{ij}\phi_j) \label{12}
\end{equation}
where ${W^{\mu\nu}_{ij}}={}g^{\mu\nu}\delta_{ij}$.
In our case, the Lagrangian is transformed according to the following procedure.
First, we introduce complex fields $h\equiv(h_1+ih_2)2^{1/2}$ and $a\equiv(a_1+ia_2)2^{1/2}$
where $h_1$, $h_2$, $a_1$,$a_2$ are identical with $h$, $a$. 
To fit into the (\ref{12}), integral by parts should be performed for the terms containing $D\phi^*$ as follows,
\begin{subequations}
\begin{align}
hQ(Da)^*={}&-DhQa^*-h(DQ)a^*\\
hA(D\varphi)^*={}&-DhA\varphi^*-h(DA)\varphi^*\\
a\tilde\varphi(D\varphi)^*={}&-Da\tilde\varphi\varphi^*-a(D\tilde\varphi)\varphi^*\\
\varphi A(D\varphi)^*={}&-D\varphi A\varphi^*-\varphi(DA)\varphi^*\label{13}
\end{align}
\end{subequations} 
Second, we replace $h^*_{\alpha\beta}P^{\alpha\beta\rho\sigma}\rightarrow h^*_{\rho\sigma}$ and $a^*_\alpha g^{\alpha\beta}\rightarrow a^*_\beta$ which are not change counter Lagrangian according to lemma in \cite{Hooft2}.
And finally, double-derivative terms are expressed in terms of $\tilde D$ which is not work on explicit field indices as follows,
\begin{subequations}
\begin{align}
\begin{split}
h^*_{\alpha\beta} D_\nu D^\nu h_{\alpha\beta}={}&h^*_{\alpha\beta}\tilde D_\nu\tilde D^\nu h_{\alpha\beta}\\
{}&+2h^*_{\alpha\beta}{\cuN^\mu}_{\alpha\beta}^{\rho\sigma}\tilde D_\mu h_{\rho\sigma}+h^*_{\alpha\beta}\cuT_{\alpha\beta}^{\rho\sigma}h_{\alpha\beta}
\end{split}\\
a^*_\alpha D_\nu D^\nu a_\alpha={}&a^*_{\alpha}\tilde D_\nu\tilde D^\nu a_{\alpha}+2a^*_{\alpha}{n^\mu}_{\alpha}^{\beta}\tilde D_\mu a_{\beta}+a^*_{\alpha}\tau_{\alpha}^{\beta}a_{\beta}\label{14}
\end{align}
\end{subequations}
where
\begin{subequations}
\begin{align}
{\cuN^\mu}_{\alpha\beta}^{\rho\sigma}={}&-2g^{\mu\lambda}\Gamma^\rho_{\lambda\alpha}\delta^\sigma_\beta\\
\cuT_{\alpha\beta}^{\rho\sigma}={}&(D_\mu\cuN^\mu+\cuN_\mu\cuN^\mu)_{\alpha\beta}^{\rho\sigma}\\
{n^\mu}_{\alpha}^{\beta}={}&-g^{\mu\lambda}\Gamma^\rho_{\lambda\alpha}\\
\tau_{\alpha}^{\beta}={}&(D_\mu n^\mu+ n_\mu n^\mu)_{\alpha}^{\beta}\label{15}
\end{align}
\end{subequations}

Applying these formula, the Lagrangian in the scalar form is obtained in terms of 10+4+1 independent complex fields $\phi_i=(h_{\mu\nu},a_\mu,\varphi)$  with
\begin{widetext}
\begin{subequations}
\begin{align}
{N^\mu}_{NG}={}&
\begin{pmatrix}
{\cuN^\mu}_{\alpha\beta}^{\rho\sigma} &  (P^{-1}\frac1 2Q^\mu)^\delta_{\alpha\beta} & \frac1 2P^{-1}C^{\alpha\beta\mu} \\
-\frac1 2g_{\gamma\lambda}Q^{\rho\sigma\mu\lambda} & {n^\mu}^{\delta}_{\gamma} & \frac1 2i\delta^\mu_\gamma{\tilde\varphi}^*\\
-\frac1 2B^{\alpha\beta\mu} & \frac1 2ig^{\mu\gamma}\tilde\varphi & -iA^\mu\label{16.1}
\end{pmatrix}
\\ M_{NG}={}&
\begin{pmatrix}
P^{-1}(X_g+X_s+X_e)+\cuT & iP^{-1}(B^{\alpha\beta\rho}-C^{\alpha\beta\rho}) & iP^{-1}C^{\alpha\beta\rho}A_{\rho} \\
-g_{\gamma\lambda}\partial_\mu Q^{\rho\sigma\mu\lambda}+ig_{\gamma\lambda}(B^{\alpha\beta\lambda}-C^{\alpha\beta\lambda})
& R^\delta_\gamma-2\delta^\delta_\gamma{\tilde\varphi}^*\tilde\varphi+\tau^\delta_\gamma & -i\tilde D_\nu{\tilde\varphi}^*+A_\nu{\tilde\varphi}^*\\
-\partial_\rho B^{\alpha\beta\rho}+iA_\rho B^{\alpha\beta\rho} & 2i\tilde D^\nu\tilde\varphi & -A_\nu A^\nu-i\partial_\nu A^\nu\label{16.2}
\end{pmatrix}
\end{align}
\end{subequations}
\end{widetext}
Since the ghost Lagrangian already has desired form, the $N^\mu_G$ and $M_G$ is written directly as follows:  
\begin{subequations}
\begin{align}
{N^\mu}_G={}&
\begin{pmatrix}
{n^\mu}_\alpha^\beta & 0 \\
-\frac1 2F_{\lambda\beta} & 0
\end{pmatrix}
\\ M_G={}&
\begin{pmatrix}
-R^\beta_\alpha+\tau^\beta_\alpha & 0 \\
-D^\lambda F_\lambda^\beta & 0
\end{pmatrix}\label{17}
\end{align}
\end{subequations}
Note that the factor $(-g)^{1/4}e_{\alpha\beta}$ is absorbed into $\phi^*$ and also $\tilde D_\nu$ is applied as non-ghost case.

With these $\{W^{\mu\nu}_{ij},M^{\mu}_{ij},N_{ij}\}$ the known results are the followings \cite{Hooft1,Hooft2}.
First, if ${W^{\mu\nu}_{ij}}={}\delta^{\mu\nu}\delta_{ij}$, it is possible to regard the propagator as $\delta_{ij}/(2\pi)^4i({k^2-i\epsilon})$
and the external vertices are corresponding to the each element of $M^{\mu}_{ij},N_{ij}$.
Although ${W^{\mu\nu}_{ij}}={}g^{\mu\nu}\delta_{ij}$ as our case, the same consideration is established by appropriate subtitution
and when we calculate in terms of  $M^{\mu}_{ij},N_{ij}$, it should be noted that there are more one-loops coming from that substitution to be considered such as  
\begin{equation}
\text{tr}{((M-D_\mu N^\mu-N_\mu N^\mu)R)},\text{tr}{(R^2)},\text{tr}{(R_{\mu\nu}R^{\mu\nu})} \nonumber 
\end{equation}
Second, the tadpole diagrams with one $M^{\mu}_{ij}$ or $N_{ij}$ are zero
and the one-loop diagrams with the product of possible combinations of $M^{\mu}_{ij},N_{ij},R_{\mu\nu},R$  by power counting is physically meaningful only in diagonal parts. 
In summary, all possible one-loop diagrams has the one of the follwoing forms,
\begin{figure}[h]
\begin{tabular}{l l}
\begin{tikzpicture}[scale=0.40]
  \def\vertices{
    \vertex[dot](a) at (0, 0){};
    \vertex (up) at (1, 1);
    \vertex (dn) at (1,-1);
    \vertex[dot](b) at (2, 0){};
  }
  \begin{feynman}
    \vertices
    \diagram* {
      (a) -- [quarter left ] (up)  -- [quarter left] (b),
      (a) -- [quarter right] (dn) -- [quarter right] (b)
    };
  \end{feynman}
\end{tikzpicture}
&
\(\displaystyle \propto \text{tr}(MM),...\) \\[0.4cm]
\begin{tikzpicture}[scale=0.40]
  \def\vertices{
    \vertex (a) at (0, 0);
    \vertex[dot] (up) at (1, 1){};
    \vertex[dot] (dn) at (1,-1){};
    \vertex[dot] (b) at (2, 0){};
  }
  \begin{feynman}
    \vertices
    \diagram* {
      (a) -- [quarter left ] (up)  -- [quarter left] (b),
      (a) -- [quarter right] (dn) -- [quarter right] (b)
    };
  \end{feynman}
\end{tikzpicture}
&
\(\displaystyle \propto \text{tr}(MN_\mu N^\mu),...\) \\[0.4cm]
\begin{tikzpicture}[scale=0.40]
  \def\vertices{
    \vertex[dot] (a) at (0, 0){};
    \vertex[dot] (up) at (1, 1){};
    \vertex[dot] (dn) at (1,-1){};
    \vertex[dot] (b) at (2, 0){};
  }
  \begin{feynman}
    \vertices
    \diagram* {
      (a) -- [quarter left ] (up)  -- [quarter left] (b),
      (a) -- [quarter right] (dn) -- [quarter right] (b)
    };
  \end{feynman}
\end{tikzpicture}
&
\(\displaystyle \propto \text{tr}(N^\mu N_\mu N^\nu N_\nu),...\) 
\end{tabular}
\caption{One-loop diagrams in Einstein-SQED system}
\end{figure}
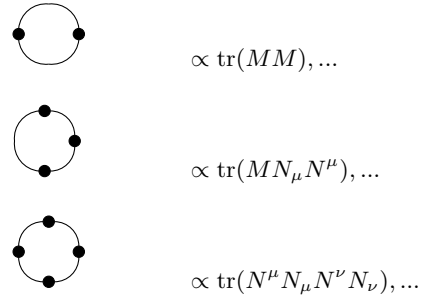\\
note that the number of the ommited legs attached at vertices can be up to 4
and when we calculate exact counter-Lagrangian, the integral parts multiplied with each trace might contain more invariants such as the tr({$\partial N\partial N$).

Unlike Einstein-Scalar or Einstein-Maxwell case, the $M^{\mu}_{ij},N_{ij}$ in Einstein-SQED system contains scalar-photon vertices with subindex $ij=23$ or $ij=32 $. 
And vertices in  Einstein-Scalar and Einstein-Maxwell systems are corrected by the amount in corresponding $M^{\mu}_{ij},N_{ij}$.
For example, the one-loop diagram with graviton-photon vertices is
\begin{figure}[h]
\feynmandiagram {
  c -- [double] a [dot,label=160:\(M_{12}\)]
  --[gluon,half left,edge label'=\(h\)] b [dot,label=20:\(M_{21}\)]
  -- [double] d,
  b  --[photon,half left,edge label'=\(a\)] a,
};
\caption{One-loop diagram with graviton-photon vertices}
\end{figure}
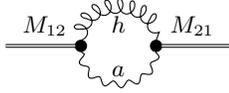\\
here, the external double lines represent the fuction of external fields. 
The counter-term corresponding to above diagram is
\begin{equation}
\Delta\Lagr = \frac1 {8\pi^2(n-4)}\frac1 4M_{12}M_{21}\label{18}
\end{equation}
From the (\ref{16.2}) the function of scalar fields $B^{\alpha\beta\rho}-C^{\alpha\beta\rho}$ should be counted in the calculation of counter-term. 

\section{Non-renormalizability}
So far, we have considered how to calculate the one-loops in Einstein-SQED system. In this section, we show the non-renormalizability of Einstein-SQED system. 
The total counter-Lagrangian is represented by the sum of all possible one-loops contribution : 
\begin{equation}
\Delta\Lagr=\frac1 \epsilon(-g)^{1/2}\{\text{tr}[\frac1 {12}Y_{\mu\nu}Y^{\mu\nu}+\frac1 2X^2+\frac1 {60}(R_{\mu\nu}R^{\mu\nu}-\frac1 3R^2)]\} \label{19}
\end{equation}
\begin{subequations}
where
\begin{equation}
Y_{\mu\nu}=\partial_\mu N_\nu-\partial_\nu N_\mu+N_\mu N_\nu-N_\nu N_\mu
\end{equation}
\begin{equation}
X=M-D_\mu N^\mu-N_\mu N^\mu-\frac1 6R\label{20}
\end{equation}
\end{subequations}
Note that the trace is to be taken over the 15 independent fields (10($h_{\mu\nu}$)+1($\varphi$)+4($a_\mu$)) for the non-ghost parts, and 5 independent fields (4($\phi^a$)+1($\chi$)) for the ghost parts. 

But we use the equation of motion and already known result rather than lengthy calculation. 
The equation of motion is obtained by requiring that the action is stationary with respect to variations.
Then the equation of motion for each the quantum fields $h_{\mu\nu},a_\mu,\varphi$ are
\begin{subequations}
\begin{align}
{}&\tilde D_\mu\tilde D^\mu\tilde\varphi=0\qquad\tilde D_\mu\tilde D^\mu{\tilde\varphi}^*=0\\
{}&D_\alpha F^{\alpha\beta}=i({\tilde\varphi}^*\tilde D^\mu\tilde\varphi-\tilde\varphi\tilde D^\mu{\tilde\varphi}^*) \\
{}&R_{\mu\nu}-\frac1 2g_{\mu\nu}R=-\frac1 2T_{\mu\nu} \label{21}
\end{align} 
\end{subequations}
where the energy-momentum tensor is
\begin{equation}
\begin{split}
T_{\mu\nu}={}&-\frac1 2(2\tilde D_\mu\tilde\varphi^* \tilde D_\nu\tilde\varphi-g_{\mu\nu}\tilde D^\alpha\tilde\varphi^* \tilde D_\alpha\tilde\varphi\\
{}&+F_{\mu\alpha}F^\alpha_\nu-\frac1 4g_{\mu\nu}F^{\alpha\beta}F_{\alpha\beta})\label{22}
\end{split}
\end{equation}
It is also possible to express $R_{\mu\nu}$ and $R$ separately by taking the trace of (\ref{21}) as follwing 
\begin{subequations}
\begin{align}
R={}&-\tilde D_\alpha\tilde\varphi^* D^\alpha\tilde\varphi-\frac1 4F_{\alpha\beta}F^{\alpha\beta}\\
R_{\mu\nu}={}&-\tilde D_\mu\tilde\varphi^* D_\nu\tilde\varphi-\frac1 2F_{\mu\alpha}F^\alpha_\nu\label{23}
\end{align}
\end{subequations} 
On the other hand, Einstein-Scalar and Einstein-Maxwell system are non-renormalizable with $\alpha R^{\mu\nu}R_{\mu\nu}+\beta R^2$ form.
Since Einstein-SQED system contains scalar-electromagnetic interaction terms that are not in these systems,  
the candidates that can remove divergent terms are $\partial F\partial \varphi$ and $\partial\partial\varphi F$. 
But these terms can not be transformed into the energy-momentum tensor form of Einstein equation because the equations are coupled. 
Hence, the $R_{\mu\nu} , R^2$ can not be eliminated even after applying the equation of motions.
Therefore the theory of the Einstein-SQED is non-renormalizable.

\section{Conclusions}
In this paper, the algorithm for one-loop diagrams of Einstein-SQED is derived in non-supersymmetric covariant theory.
By adding scalar field to Einstein-Maxwell fields, there are more possible counter-Lagrangian terms, but these do not remove the divergent terms.
On the other hand, supersymmetry provides the physical principle to add other fields, namely supersymmetric partners. 
In that case, there are miraculous cancellations in loop calculations. 

\end{document}